\documentclass[aps,prl,twocolumn,10pt,superscriptaddress,showpacs]{revtex4-1}

\usepackage{graphicx}
\usepackage{hyperref}
\usepackage{url}
\usepackage{amsmath}
\usepackage{verbatim}
\usepackage{color}

\def\beq {\begin{equation}}
\def\eeq {\end{equation}}

\def\bfr {\mathbf{r}}

\def\bfq {\mathbf{q}}

\date{\today}

\begin{document}
\title{Exciton band structure in two-dimensional materials}

\newcommand{\lsi}{Laboratoire des Solides Irradi\'es, \'Ecole Polytechnique, CNRS, CEA,  Universit\'e Paris-Saclay, F-91128 Palaiseau, France}
\newcommand{\etsf}{European Theoretical Spectroscopy Facility (ETSF)}
\newcommand{\soleil}{Synchrotron SOLEIL, L'Orme des Merisiers, Saint-Aubin, BP 48, F-91192 Gif-sur-Yvette, France}
\newcommand{\london}{Department of Physics, King's College London, London WC2R 2LS, United Kingdom}

\author{Pierluigi Cudazzo}
\affiliation{\lsi}
\affiliation{\etsf}

\author{Lorenzo Sponza}
\affiliation{\london}

\author{Christine Giorgetti}
\affiliation{\lsi}
\affiliation{\etsf}

\author{Lucia Reining}
\affiliation{\lsi}
\affiliation{\etsf}

\author{Francesco Sottile}
\affiliation{\lsi}
\affiliation{\etsf}

\author{Matteo Gatti}
\affiliation{\lsi}
\affiliation{\etsf}
\affiliation{\soleil}

\begin{abstract}
Low-dimensional materials differ from their bulk counterpart in many respects. In particular, the screening of the Coulomb interaction is strongly reduced, which can have important consequences such as the significant increase of exciton binding energies. In bulk materials the binding energy 
is used as an indicator in optical spectra to distinguish different kinds of excitons, 
but this is not possible in low-dimensional materials, where the binding energy is large and comparable in size for excitons of very 
different localization.
Here we demonstrate  that the exciton band structure, which can be  accessed experimentally, instead provides  a powerful way to  identify the exciton character.
By comparing the {\it ab initio} solution of the many-body Bethe-Salpeter equation for graphane and  single-layer  hexagonal BN,
we draw a general picture of the exciton dispersion in two-dimensional materials, highlighting the different role played by the exchange electron-hole interaction
and by the electronic band structure. Our interpretation is substantiated by a prediction for phosphorene.

\end{abstract}


\maketitle

One of the most intriguing features of two-dimensional (2D) materials is the emergence of fundamentally distinct physical properties from those of their bulk counterparts.  
The unique properties exhibited by 2D materials are associated with the evolution of the electronic band structure as the single-layer limit is approached.
A prominent example is graphene, where the linear band-dispersion at the K point
gives rise to novel phenomena \cite{Novoselov2005,CastroNieto2009} such as the anomalous integer quantum Hall effect at room temperature \cite{Zhang2005,Novoselov2007}.
More recently, in transition-metal dichalcogenides (TMD) strong photoluminescence has been demonstrated 
occurring concomitantly with the crossover between indirect and direct band gap \cite{Mak2010,Splendiani2010}, when the dimensionality is reduced from the bulk to the monolayer. 
These materials display novel excitonic properties, related to 
efficient control of  valley and spin occupation by optical helicity \cite{Xiao2012,Mak2012,Zang2012,Cao2012}. 
These findings are  driving considerable  renewed interest in the excitonic physics of low-dimensional materials,
both from the point of view of fundamental physics and for technological applications \cite{Wang2012,Xu2014,Yu2014,Li2014,MacNeill2015,Srivastava2015,Aivazian2015,Xia2014,Liu2015}.
It is therefore crucial to be able to distinguish features that are specific for certain materials from others that characterize 2D systems in general, and to obtain a deep understanding that allows one to make predictions and eventually design materials with desired properties. 

A popular and non-destructive approach to study low-dimensional systems is optical experiments. However, although dimensionality changes
lead to modifications in the band structure, in many situations
optical absorption spectra (at vanishing in-plane momentum transfer $\bfq\rightarrow0$) remain unaltered 
due to cancellation effects \cite{Wirtz2006}.
This occurs in a large variety of systems: for example, for the so-called A and B excitons in insulating TMDs \cite{Molina2013}, or the tightly bound exciton in hexagonal (h)-BN \cite{Wirtz2006}. 
On the contrary, the exciton binding energies (EBE) in 2D systems are in general much larger than in their 3D counterparts, especially 
for materials where the 3D EBEs are small \cite{Wirtz2006,Molina2013,Cudazzo2010,Luo2011,Hsueh2011,Qiu2013,Chernikov2014,He2014,Hanbicki2015,Wang2014}.
Therefore EBEs are similar in very different 2D systems and, contrarily to the situation in 3D, they cannot be used to
distinguish excitons of different character. One might even wonder whether  there \emph{are}
excitons of significantly different character in 2D, and in particular whether they show a significantly different degree of charge localization.
If this is the case, the question arises how these excitons could be distinguished, since the EBE is not discriminating. 
In the present work we demonstrate that there are different classes of excitons in 2D, similarly to the 3D case,
and that the clue to detect and understand their character is to 
go beyond the limiting case of optical absorption, exploring excitonic spectra over a wide range of momentum transfer $\bfq$. 
Since the exciton dispersion can be measured, this offers 
experimental access to the characterization of excitons, also in low dimensional materials.

Numerous studies of the dispersion of excitations exist for plasmons. They directly reveal dimensionality effects.
For example, the $\pi$ plasmon in graphite shows a quadratic  $\bfq$-dispersion  \cite{Zeppenfeld1971,Marinopoulos2004,Wachsmuth2014}, 
whereas in graphene it is quasilinear \cite{Kramberger2008,Wachsmuth2013,Wachsmuth2014,Liou2015}. 
In metallic TMDs, the slope of the intraband-plasmon dispersion is negative in the bulk \cite{VanWezel2011,Cudazzo2012,Faraggi2012}, 
but positive in 2D \cite{Cudazzo2013}.
The study of the $\bfq$-dispersion of elementary excitations 
is therefore a key to understand the effects of the dimensionality on the electronic properties. 
However, contrarily to the plasmon dispersion, exciton dispersion in 2D materials is a subject that to the best of our knowledge has never been investigated.
Indeed, few calculations concerning exciton dispersion exist to date, even in 3D \cite{Soininen2000,Abbamonte2008,Lee2013,Gatti2013,Cudazzo2013b}. This contrasts with a strong need, since 
the understanding of the mechanisms for the propagation of excitons, and their spatial localization, are
of crucial relevance, e.g. for all applications involving light harvesting and the transport of the excitation energy. 
Our calculations and analysis fill therefore an important gap in several respects.

We consider three representative insulating 2D materials, namely graphane (i.e. hydrogenated graphene) \cite{Elias2009}, phosphorene \cite{Liu2014} and a single layer of h-BN  \cite{Novoselov2005b,Jin2009,Alem2009}, and
compare their exciton band structure  $E^\lambda_\bfq$, obtained from the {\it ab initio} solution of the Bethe-Salpeter equation (BSE) as a function of the exciton wavevector $\bfq$ as described in Ref. \cite{Gatti2013}. 
The BSE can be cast into an effective two-particle Schr\"odinger  equation 
for the  wavefunction $\Psi^\lambda_\bfq(\bfr_h,\bfr_e)$ of the electron-hole (e-h) pair: 
$H_{\textrm{exc}} \Psi^\lambda_\bfq  = E^\lambda_\bfq\Psi^\lambda_\bfq$ \cite{Onida2002}.  Within the GW approximation \cite{Hedin1965} to the BSE, 
the excitonic Hamiltonian $H_{\textrm{exc}}=H_{\textrm{e}}+H_{\textrm{h}}+H_{\textrm{e-h}}$
is the sum of the independent propagations (i.e. hoppings) $H_{\textrm{e}}$ and $H_{\textrm{h}}$ of the electron and the hole (which 
derive from the GW quasiparticle (QP) band structure) and the e-h interaction $H_{\textrm{e-h}}$,
which includes the exchange electron-hole repulsion due to the bare Coulomb interaction $v_c$,
and the direct electron-hole attraction due to the statically screened Coulomb interaction $W$.
In our calculations, as discussed in the supplementary material \cite{suppmat}, we have adopted a supercell approach, using a truncation of $v_c$  \cite{IsmailBeigi2006} to prevent interactions between periodic copies. Moreover, we have
avoided divergences of Coulomb integrals in low dimensions \cite{Marini2009} by means of a 2D analytical integration that efficiently removes the Coulomb singularity \cite{suppmat,Sponzaphd}. The solution of the BSE is used to construct the macroscopic frequency- and wave vector dependent 
dielectric function $\epsilon_M(\bfq,\omega)$, from which spectra are obtained.

\begin{figure}
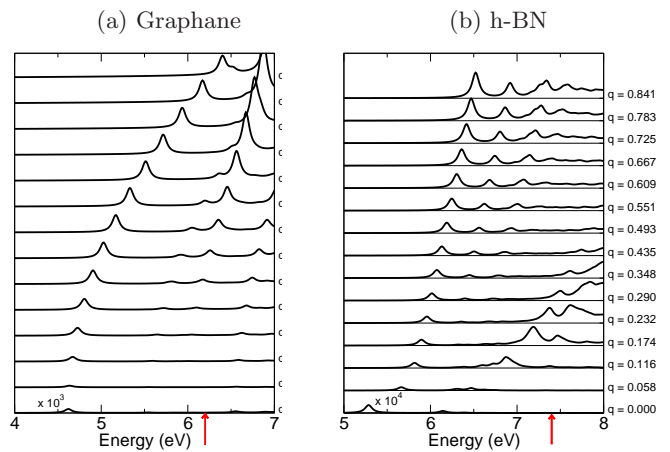

\begin{tabular}{c c}
(a) Graphane & (b) h-BN \\[0.2cm]
\includegraphics[width=0.49\columnwidth]{fig1.eps}
& \includegraphics[width=0.49\columnwidth]{fig2.eps}
\end{tabular}
\caption{${\rm Im}\, \epsilon_M(\bfq,\omega)$ of  (a) graphane and  (b) h-BN calculated for several values of $\bfq$ (in \AA$^{-1}$) along ${\Gamma}M$. For sake of clarity, each spectrum has been multiplied by $q^2$.
The red arrows mark the QP band gap.} 
\label{fig1}
\end{figure}

Fig. \ref{fig1} shows ${\rm Im}\, \epsilon_M(\bfq,\omega)$ for graphane and h-BN for different $\bfq$ along the ${\Gamma}M$ direction. 
In agreement with previous calculations at $\bfq\rightarrow0$ \cite{Wirtz2006,Cudazzo2010}, both materials display exciton peaks inside the QP gap which is marked by the red arrows in Fig. \ref{fig1}: 
at 4.6 eV in graphane and a prominent feature at 5.3 eV in h-BN.
In both cases the lowest-energy peak in the spectrum for $\bfq\rightarrow0$ is related to two degenerate bound excitons involving e-h pairs of the top valence and bottom conduction bands \cite{suppmat}: 
only one  is visible along this  direction while the other one is dark.
At finite $\bfq$  this exciton degeneracy is removed. However, in both materials only one peak is visible in the spectra since the other exciton remains dark also for $\bfq\neq0$. Interestingly, in both systems new features appear at large $\bfq$. In particular, in graphane the peak at about 6.5 eV is related to higher energy  interband transitions not visible at $\bfq\rightarrow0$, while in h-BN the series of peaks between 6.5 and 7.5 eV is related to the transitions from the top-valence to the bottom-conduction bands. In the following we will focus on the region of onset of the spectra.

\begin{figure}
\includegraphics[width=1.0\columnwidth]{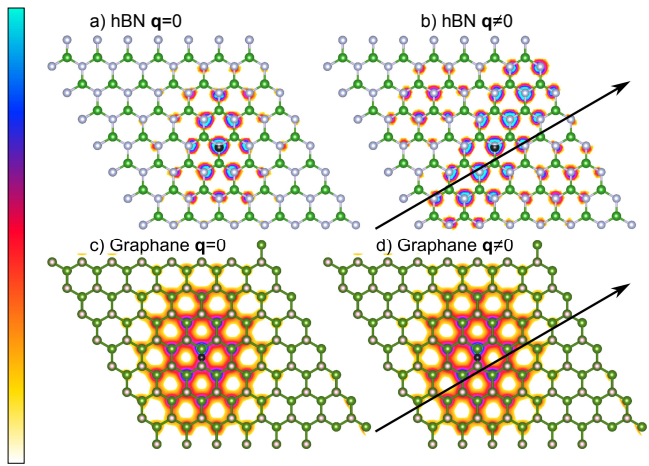} 
\caption{Electronic distribution of the wavefunction $\Psi^\lambda_{\bfq}$ of the lowest-energy exciton of graphane and h-BN
for fixed position of the hole (black sphere) calculated for $\bfq\rightarrow0$ (left) and 0.4 \AA$^{-1}$ (right). 
The  black arrow is the  direction of $\bfq \neq 0$.
White and green spheres represent N and B atoms in h-BN and H and C atoms in graphane.} \label{fig2}
\end{figure}

In Fig. \ref{fig2} we compare the electronic distribution of the wavefunctions $\Psi^\lambda_{\bfq}(\bfr_h,\bfr_e)$  for $\bfq\rightarrow0$ and $\bfq=0.4$ \AA$^{-1}$ of the lowest-energy excitons of graphane and h-BN,
fixing the position of the hole where the valence electron wavefunction is mostly localised.
This allows one to infer the spatial extension  of the exciton from the solution of the BSE, as shown in several 2D materials for $\mathbf{q}\rightarrow0$ \cite{Molina2013,Cudazzo2010,Luo2011,Hsueh2011,Qiu2013,suppmat}.
In graphane at $\bfq\rightarrow0$  the electronic charge, for a hole fixed on a C-C bond, is distributed on both C and H atoms and is delocalized over several unit cells \cite{Cudazzo2010}. 
On the contrary, in h-BN it is strongly localized around the hole, which is placed on the N atom, with a small contribution coming from the nearest-neighbour N atoms.
In h-BN  the exciton can be hence interpreted as an excitation of a ``super atom'' encompassing both N and B orbitals. 
It is localized on N sites due to the strong excitonic effects (for the analogous case of LiF in 3D, see \cite{Abbamonte2008}).

The EBE  at $\bfq\rightarrow0$ in the two systems is similar: 1.6 eV in graphane and 2.1 eV in h-BN. This is in
seeming contrast to the large difference in the exciton wavefunction. It illustrates the fact that
in low dimensions the value of the EBE measured in the absorption spectra at $\bfq\rightarrow0$ does not directly give  information about the nature of the exciton. On the contrary in the 3D case the spatial localization of excitons correlates directly with their binding energy. This allows one to classify excitons in 3D as 
localized Frenkel excitons with EBEs of the order of several eV, and delocalized Wannier-Mott excitons
with EBEs of tens of meV
\cite{Knox1963,Bassani1975}. This can be understood, since in 3D semiconductors screening is strong for large distances between electron and hole,
which reduces the binding energy of Wannier excitons.
In 2D semiconductors the macroscopic screening is much weaker
\cite{Keldysh1979,Cudazzo2011}, resulting in a strong increase of their EBE. 
At short distances instead, screening is always weak in both 2D and 3D, and therefore it does  
not play a crucial role for the EBE of Frenkel excitons, where electron and hole are close together.
Therefore, EBEs of Frenkel excitons are
less sensitive to the dimensionality, and similar to the ones of Wannier excitons in 2D.

\begin{figure}
\vspace{1cm}
\includegraphics[width=1.0\columnwidth]{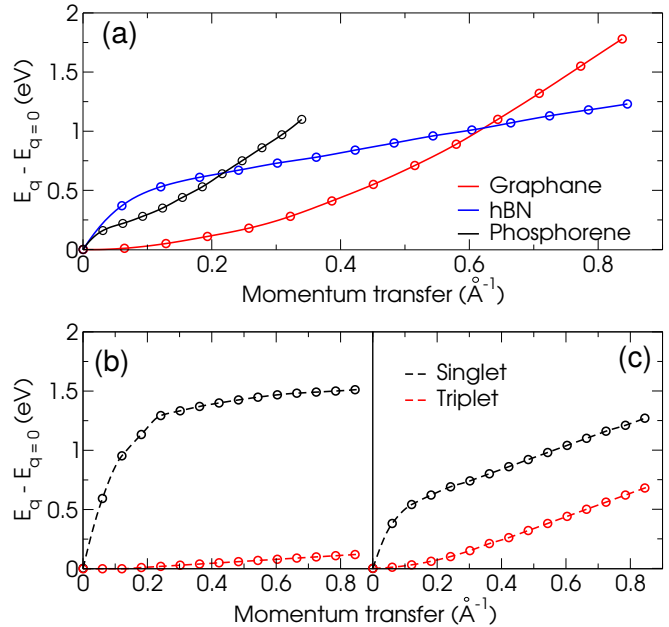}
\caption{(a) Calculated dispersion of the lowest-energy bright excitons in graphane and h-BN along ${\Gamma}M$ and phosphorene along ${\Gamma}X$.
(b) Singlet and triplet exciton dispersion in h-BN in a two-band model with flat bands and (c) with the real electronic band structure.} 
\label{fig3}
\end{figure}

In the following, we show that it is nevertheless possible to distinguish different excitons in 2D: the clue is to look at non-vanishing
wave vector $\bfq$. This is illustrated by the wavefunctions in 
Fig. \ref{fig2}. 
As $\bfq$ increases, in h-BN the electronic charge becomes anisotropically delocalized with respect to the position of the hole, 
but in graphane it is independent of $\bfq$. 
This striking result suggests that the exciton band structure $E^\lambda_\bfq$, that can also be obtained from experiment,
should be analyzed more in detail. 
Fig. \ref{fig3}(a) shows the calculated band structure of the lowest-energy bright excitons in h-BN and graphane.
Its curvature can be classified in two different categories.
In h-BN  the dispersion is linear for small $\bfq$ and the curve flattens at large $\bfq$.
In graphane instead $E^\lambda_\bfq = E_{\bfq=0}^\lambda + \alpha q^2$.
The figure also shows the exciton dispersion of phosphorene: it behaves similar to graphane with a parabolic dispersion excluding a very restricted range around $\mathbf{q}=0$ \cite{suppmat}.

By a parabolic fit of the exciton dispersion of graphane \cite{suppmat}, the effective mass $m_{\textrm{exc}}$ of the dark and bright exciton can be estimated to be 1.6 and 1.8, respectively. Both values match well the sum $m_e+m_h=1.3$ of the effective electron mass $m_e$ and  the average value $m_h$ of the effective masses of the heavy and light holes at the $\Gamma$ point in the electronic band structure.
This result evidences the Wannier character of the exciton in graphane: the dispersion is set by the center-of-mass motion of the e-h pair displaying a free-particle-like behavior (the same holds  for phosphorene \cite{suppmat}).
This also explains why the exciton wavefunction in graphane is independent of $\bfq$ (see Fig. \ref{fig2}). 
When the electron and the hole separately behave as free particles moving in a homogeneous dielectric material (i.e. in the Wannier model), it is possible to find a canonical transformation that decouples the center-of-mass motion from the relative motion of the e-h pair \cite{Wannier1937}. 
In this case the exciton state
factorizes in the product of the center-of-mass wavefunction, which propagates as a free particle with momentum $\bfq$, and an e-h correlation  function that is independent of $\bfq$. The exciton wavefunction thus depends on $\bfq$  only through a phase factor and does not change its shape as $\bfq$ increases. 

In order to rationalize the exciton band dispersion in h-BN, we solve the {\it ab initio} BSE as above but considering only transitions between the top-valence and the bottom-conduction bands in which 
we artificially set the dispersion to 0. This corresponds to neglecting the electron and hole hopping terms \cite{suppmat}.
This choice is suggested by the observation that in h-BN the lowest excited state involves e-h pairs belonging to the $MK$ line of the first Brillouin zone, where the bands have a weak dispersion \cite{suppmat}.  
In such a simplified situation with only two flat bands and assuming a negligible overlap between wavefunctions localized on different atomic sites, the $\bfq$-dispersion is that of a pure Frenkel exciton, given by \cite{Cudazzo2013b,Cudazzo2015,suppmat}:
\beq
E^{\textrm{FR}}_{\bfq} = {\Delta}E+\mathcal{I}(\mathbf{q})-\mathcal{W},
\label{E_FR}
\eeq
where ${\Delta}E$ is the energy difference between the two flat bands, 
$\mathcal{W}$ is the on-site term of the direct e-h attraction $W$, and 
 $\mathcal{I}(\bfq)$, which is  the only term that can induce a dispersion of the exciton energy, is
the excitation-transfer interaction \cite{Bassani1975} related to the exchange e-h repulsion.

Fig. \ref{fig3}(b) displays the lowest-exciton dispersion obtained from the BSE considering the two flat bands in h-BN  for both the singlet and triplet channels.
We find that the triplet exciton has a negligible dispersion, following the behavior described by  Eq. \eqref{E_FR}, since for the triplet the exchange e-h interaction is absent and thus $\mathcal{I}(\mathbf{q})=0$.
This demonstrates that in absence of hopping the lowest excited state of h-BN is a pure Frenkel exciton. 
In the singlet channel the exciton has a linear dispersion for $\bfq\rightarrow0$ and reaches a constant value at large $\bfq$. 
This behavior is due to the $\bfq$-dependence of the exchange e-h repulsion, which at small $\bfq$ can be described in terms of a dipole-dipole interaction.
In this regime, the linearity of $\mathcal{I}(\bfq)$ is in contrast with 3D bulk insulators,  where it is generally characterized by a quadratic $\bfq$-dispersion \cite{Cohen1955}.

To better understand the effect of the dimensionality on the exchange e-h interaction we consider a quasi-2D dielectric of finite thickness $d$ along the $z$ axis normal to the plane. 
We assume that the electronic orbitals $\psi(\mathbf{r}$) can be factorized in an in-plane $\phi(\mathbf{\rho})$ and out-of-plane $\chi(z)$ components:  
$\psi(\mathbf{r})=\phi(\mathbf{\rho})\chi(z)$. 
With the simplest form $\chi(z)=1/\sqrt{d}$ for $|z|<d/2$ and $\chi(z)=0$ for $|z|>d/2$,  i.e. $\chi$ constant inside the 2D slab and zero elsewhere, 
the exchange e-h interaction gives rise  \cite{suppmat} to the following dispersion (where we introduce the dipole matrix element $\boldsymbol{\mu}_{vc}$ related to the valence-conduction transition):
\begin{equation}
\mathcal{I}(\mathbf{q})=\frac{4\pi}{d}(\hat{\mathbf{q}}\cdot\boldsymbol{\mu}_{cv})(\hat{\mathbf{q}}\cdot\boldsymbol{\mu}_{vc})\left[1-\frac{1}{qd}(1-e^{-qd})\right],
\end{equation}
which is linear for $q \ll 1/d$ and becomes a constant for $q \gg 1/d$ \cite{suppmat}. 
Alternatively, the linear behavior at small $\bfq$ can be obtained using a 2D Coulomb potential in the evaluation of the exchange e-h interaction \cite{Hambachphd}. This means that in the optical limit $\bfq\rightarrow0$
the system behaves as a strictly 2D dielectric. At large $\bfq$, on the other hand, when $1/q$ becomes comparable with $d$, finite-thickness effects become important due to the 3D nature of the Coulomb interaction. 
The system does not behave anymore as an infinite 2D dielectric, but rather as a 3D finite system
\footnote{A similar effect of the finite thickness can be found also on the intraband plasmon dispersion of 2D metals \protect\cite{Ritchie1957,Ferrel1958}.}. 
As a consequence, due to the absence of the hopping, at large $\bfq$ the exciton stops dispersing. 
This means that the exciton band dispersion displayed in Fig. \ref{fig3}(b) is not materials specific but general, being strictly related to the confinement of the electronic charge in a 2D slab and its effect on the e-h exchange. 
This interpretation is further confirmed by the fact that the dark exciton along ${\Gamma}M$  does not disperse \cite{suppmat} (with $\boldsymbol{\mu}_{vc}=0$ one has $\mathcal{I}(\mathbf{q})=0$).

When we solve the BSE with the real dispersion of the top-valence and bottom-conduction bands [see Fig. \ref{fig3}(c)], the singlet exciton
is the same as from the full calculation that takes into account all the bands (see Fig. \ref{fig3}(a)). 
This confirms that the exciton in h-BN mainly originates from transitions between that pair of bands.
The hopping, i.e. the dispersion of the electronic band structure,
modifies the dispersion of the singlet exciton with respect to the flat-band limit.
Moreover, by taking into account the band dispersion, also the triplet exciton  in Fig. \ref{fig3}(c) acquires a dispersion. 
The correspondence of the triplet exciton and electronic 
band dispersions \cite{suppmat} suggests that the contribution coming from the hopping term is of second or higher order in $\bfq$,
as expected also from a tight-binding description. 
This also implies that the effect of the hopping on the singlet exciton is negligible in the optical limit, where the dispersion is dominated by the exchange e-h interaction, which is linear in $\mathbf{q}$. 
On the other hand, at large $\bfq$  where $\mathcal{I}(\bfq)$ becomes a constant, the dispersion is set only by the hopping in both singlet and triplet channels. 
Indeed, in Fig. \ref{fig3}(c) we can observe that at large $\bfq$ triplet and singlet excitons have the same dispersion.

The mechanism at the basis of the spatial delocalization of the exciton in h-BN  can be also analogously understood as the effect of the coupling, induced by the hopping, between the pure Frenkel and other more delocalised excitons \cite{Cudazzo2013b,Cudazzo2015,suppmat} (this is the reason why the exciton in h-BN is delocalised also on nearest-neighbour sites, see Fig. \ref{fig2}).
In particular we can distinguish two different mechanisms. 
The first one is related to the $\bfq$ dependence of the  exciton energy through the exchange e-h interaction [see Eq. \eqref{E_FR}]. 
As $\bfq$ increases the energy of the pure Frenkel exciton gets closer to the other delocalised excitons (see Fig. \ref{fig1})
enhancing the coupling between them. This causes an isotropic delocalization of the exciton wavefunction. 
The second effect is related to the explicit  $\bfq$-dependence of the hopping term. 
As $\bfq$ increases,
this gives rise to the anisotropic delocalization of the exciton wavefunction in h-BN \cite{suppmat} that is displayed in Fig. \ref{fig2}.

In conclusion, we have shown that in 2D materials excitons of different character cannot be distinguished by their binding energy,
but by their dependence on the exciton wave vector $\bfq$. Striking differences between Wannier and Frenkel-like excitons are observed in  
the exciton wavefunction $\Psi^\lambda_{\bfq}$, which is obtained from the numerical solution of the BSE. These differences are also present 
in
the exciton band structure, which
can be readily accessed experimentally by momentum-resolved electron energy loss spectroscopy (EELS) \cite{Everton2011,Wachsmuthphd} or resonant inelastic X ray spectroscopy (RIXS) \cite{Schuelke2007,Ament2011}.
Therefore, the present work suggests to use the exciton band structure as a direct and powerful indicator
for the exciton character in 2D systems and the relative importance of the different e-h interactions at play in the materials. 
Our conclusions are supported by numerical results for
three prototypical 2D semiconductors.
An analysis based on general arguments shows the the findings are not materials specific, but generally valid for all 2D materials.

This research was supported by the European Research Council (ERC Grant Agreement n. 320971)
and by a Marie Curie FP7 Integration Grant within the 7th European Union Framework Programme.
Computational time was granted by GENCI (Project No. 544).

\bibliographystyle{apsrev4-1}

%

\end{document}